\documentstyle [12pt]{article}
\begin{document}
\begin{center}
{\Large\bf Black Hole thermodynamics without a black hole?}
\end{center}
\vspace{5mm}
\begin{center}
{\large\bf Victor Berezin}
\end{center}
\vspace {3mm}
\begin{center}
Institute for Nuclear Research, Russian Academy of Sciences,\\ 
60th October Anniversary prospect, 7a,Moscow, Russia \\
berezin@ms2.inr.ac.ru \\ and \\ Max Plank Institute for Gravitational Physics,
\\ Albert Einstein Institute, Am Muhlenberg, 1, Golm, Germany
\end{center}
\vspace{20mm}
\begin{center}
{\large\bf Abstract}
\vskip0.5cm
\end{center}
In the present paper we consider, using our earlier results, the process 
of quantum gravitational collapse and argue that there exists the final 
quantum state when the collapse stops. This state, which can be called 
the ``no-memory state'', reminds the final ``no-hair state'' of the 
classical gravitational collapse. Translating the ``no-memory state'' into 
classical language we construct the classical analogue of quantum black hole 
and show that such a model has a topological temperature which equals 
exactly the Hawking's temperature. Assuming for the entropy the 
Bekenstein-Hawking value we develop the local thermodynamics for our model 
and show that the entropy is naturally quantized with the equidistant 
spectrum $S + \gamma_0 N$. Our model allows, in principle, to calculate 
the value of $\gamma_0$. In the simplest case, considered here, we obtain 
$\gamma_0 = \ln{2}$.
\newpage
\begin{center}
{\large\bf Preliminaries}	
\end{center}
\vskip0.5cm
\noindent
In 1972 J.D.Bekenstein observed the striking resemblance of the 
Schwarzschild black hole mechanics with the first and second laws of 
thermodynamics \cite{1}. He presented very serious physical arguments 
that the Schwarzschild black hole should be ascribed by a certain amount 
of entropy which is proportional to the event horizon area. In 1973 
J.M.Bardeen, B.Carter and S.W.Hawking extended this idea and proved the 
four laws of thermodynamics for the general class of Kerr-Newman black 
hole \cite{2}, the role of the temperature being played by the surface 
gravity (up to some numerical factor), which is constant along the event 
horizon. And only  after discovering by S.W.Hawking the black hole 
evaporation \cite{3} this analogy became the real physical phenomenon. 
It appeared that the spectrum of such a radiation is Planckian with the 
temperature $T_H = \frac{\kappa}{2 \pi}$, where $\kappa$ is the surface 
gravity. It follows then that the black hole entropy is exactly one 
fourth of the dimensionless horizon area, 
\begin{equation}
\label{S}
S = \frac{1}{4} \frac{A}{l_{Pl}^2}, 
\end{equation}
where $l_{Pl} = \sqrt{\frac{\hbar G}{c^3}} \sim 10^{-33} cm$ is the Planckian 
length ($\hbar$ is the Planck constant, $c$ is the speed of light, and $G$ 
is the Newtonian gravitational constant). We use the units $\hbar = c = k 
= 1$ ($k$ is the Boltzmann constant), so $l_{Pl} = \sqrt{G}$ and the 
Planckian mass is $m_{Pl} = \sqrt{\frac{\hbar c}{G}} = 1/\sqrt{G} 
\sim 10^{-5} gr$. 

The nature of such a radiation and its black body spectrum lies in the 
nontrivial causal structure of the space-times containing black holes. 
The crucial point is the existence of the horizons. The same takes place 
in the Rindler spacetime which is actually the part of the flat Minkowski 
spacetime with the event horizon. The Rindler's observer experiences 
a constant acceleration $a$ and ``sees'' a thermal bath with the 
temperature $\frac{a}{2 \pi}$ which is called the Unruh's temperature 
\cite{4}. The Hawking's temperature $T_{H}$ is just the Unruh's 
temperature $T_{U}$ at the event horizon.

The quantum nature of the radiation implies the quantization of the black 
hole mass. The first attempt was made by J.D.Bekenstein \cite{5}. He 
noticed that the horizon area of non-extremal black holes behaves as a 
classical adiabatic invariant. The Bohr-Sommerfeld quantization rule 
then predicts the equidistant discrete spectrum for the horizon area 
and, thus, for the black hole entropy. The gedanken experiments show 
that the minimal increase in the horizon area in the process of 
capturing the neutral \cite{6} or electrically charged \cite{7} 
particle is approximately equals to
\begin{equation}
\label{A}
\Delta A_{min} \approx 4 l_{Pl}^2 .
\end{equation}
This suggests for the black hole entropy (provided we accept the relation 
(\ref{S})) 
\begin{equation}
\label{sbh}
S_{BH} = \gamma_0 N,\qquad N = 1,2,...
\end{equation}
where $\gamma_0$ is of order of unity. In their famous work on the 
black hole spectroscopy J.D.Bekenstein and V.F.Mukhanov \cite{8} related 
the black hole entropy to the number of microstates $g_n$ that corresponds 
to a particular external macrostate through the well known formula in 
statistical physics, $g_n = \exp{[S_{BH}(n)]}$, e.i., $g_n$ is the 
degeneracy of the $nth$ area eigenvalue. Since $g_n$ should be an integer 
they deduce that 
\begin{equation}
\label{g}
\gamma_0 = \ln{k}, \qquad k = 2,3,...
\end{equation}
In the spirit of the information theory and  ``It from Bit''-idea 
by J.A.Wheeler the value of $\ln{2}$ seems the most suitable one. The 
equidistant area spectrum was also derived from the some symmetry 
principles \cite{8a,8b,8c}.

The confirmation of microscopical statistical nature of the black hole 
entropy came from the string theory. A.Strominger and C.Vafa \cite{9} 
were the first who counted directly the degeneracy of the horizon microstates 
in the special case of 5-dimensional extremal black hole and showed that 
the relation (\ref{S}) is exact. A review of further progress can be found 
in \cite{10}. 

The very natural way of counting the number of microscopic states is 
provided by the Loop Quantum Gravity (see \cite{11} for a recent review). 
In this rather new approach to canonical quantization of gravity the 
area operator has a discrete spectrum. Such an operator can be represented 
by a spin network puncturing a surface. The procedure of counting the 
surface states at the horizon was developed by K.Krasnov \cite{12} and 
applied to calculating the black hole entropy by A.Ashtekar et al. 
\cite{13,14}. The net result is that the entropy of the spherically 
symmetric black hole equals  
\begin{equation}
\label{LS}
S_{BH} = N \ln{(2 j_{min} + 1)} ,
\end{equation}
where $j_{min}$ is the minimal (nonzero) spin value depending on the 
underlying symmetry group. In the conventional Loop Quantum Gravity this 
is the $SU(2)$ group, thus, $j_{min} = \frac{1}{2}$, and for $\gamma_0$ 
(Eqn.(\ref{g})) we have $\gamma_0 = \ln{2}$. For the $SO(3)$ group 
$j_{min} = 1$, and $\gamma_0 = \ln{3}$. For the horizon area Loop Quantum 
Gravity gives in this case the value 
\begin{equation}
\label{LA}
A_h = 8 \pi \gamma \sqrt{j_{min} (j_{min} + 1)} N l_{Pl}^2 ,
\end{equation}
where $\gamma$ is the so-called Immirzi ambiguity parameter \cite{15}. 
It equals $\frac{\ln{2}}{\sqrt{3} \pi} (\approx 0.12738402)$ for 
$j_{min} = \frac{1}{2}$ and $\frac{\ln{3}}{2 \sqrt{2} \pi} 
(\approx 0.12363732)$ for $j_{min} = 1$, provided $S_{BH} = 
\frac{1}{4} \frac{A_h}{l_{Pl}^2}$. Thus, Loop Quantum Gravity gives us 
almost unique (up to the choice of $j_{min}$ and, of course for large N) 
quantum spectrum for the black hole entropy, but the horizon ares and, 
hence, the black hole mass spectra depend on the choice of the Immirzi 
parameter.

The recent progress in this subject is connected to the so-called 
quasi-normal modes of the Schwarzschild black hole. It is known for a long 
time that the decay of black hole perturbations is dominated at late times 
by a set of damped oscillations (see, e.g. \cite{16}). It was shown that 
for the frequencies $\omega$ with large imaginary part, the real part 
becomes equally spaced, and 
\begin{equation}
\label{omega}
m \omega = 0.04371235 + \frac{i}{4} (n + \frac{1}{2}) ,
\end{equation}
where $m$ is the black hole mass \cite{17,18}. 
S.Hod noticed \cite{19} that the real part of $\omega$ can be written as 
(and it was later proven analytically \cite{20}) 
\begin{equation}
\label{qnm}
\omega_{QNM} = \frac{\ln{3}}{8 \pi m} .
\end{equation}
The Bohr's correspondence principle requires $dm = \omega_{QNM}$, and 
for the entropy we obtain $S_{BH} = N \ln{3}$.

In all the above mentioned approaches to quantizing the black hole area 
(or the entropy content) the event horizon is considered essentially 
classical. But, in quantum theory there are no trajectories, no geodesics 
to probe the spacetime geometry, so, the very notion of the event 
horizon is not defines. Therefore, there exists no definition of what 
a quantum black hole is. 

To overcome this difficulty we construct some very simple classical model 
(namely, the self-gravitating dust shell), then quantize it using 
minisuperspace formalism and try to extract some physical information 
\cite{21,22,23,24}. In the present paper we give a 
short outline of the classical model, the quantization procedure and the 
resulting mass spectra. Then we argue that the very process of quantum 
gravitational collapse gives rise to the increase of entropy (which is 
initially zero). The final stage of the quantum collapse is a special 
``no-memory'' quantum state that resembles the black hole ``no-hair'' 
feature. We named it ``a quantum black hole''. At the end of the paper 
we show that it is possible to construct the classical analogue of such 
a quantum state. This classical analogue possesses a topological temperature 
which coincides exactly with the Hawking's temperature for the 
Schwarzschild black hole. We give also a complete thermodynamical 
description of the model, derive the equidistant area (and entropy) 
spectrum and show how the entropy units can, in principle, be calculated.   

\newpage
\begin{center}
{\large\bf Classical Model}
\end{center}
\vskip0.5cm
\noindent
Everybody knows what the classical black hole is. In short,
black hole is a region of a space-time manifold beyond an
event horizon. In turn, an event horizon is a null surface
that separates the region from which null geodesics can
escape to infinity and that one from which they cannot.
It is important to stress that the notion of the of
the event horizon is global, it requires knowledge of both
past 
and future histories. In classical physics we have
trajectories of 
particles, we have geodesics, so, everything can be, in
principle, 
calculated. In quantum physics there are no trajectories and
the 
event horizon can not be defined. Thus, we have to seek for 
quite a different definition of a quantum black hole. Till
now 
we have no consistent theory of quantum gravity. All this
forces 
us to start with considering some models. The simpler, the 
better. 

The simplest is the so-called Schwarzschild eternal black hole. 
Its geometry is a geometry of non-traversable wormhole. 
There are two asymptotically flat regions at spatial 
infinities connected by the Einstein-Rosen bridge. The 
gravitating source is concentrated at two space-like 
singular surfaces or zero radius. Two sides of the 
Einstein-Rosen bridge are causally disconnected and 
separated by event horizons. The narrowest part of the 
bridge is called a throat, its size is the size of the 
horizon. Eternal black holes are parameterized by 
total (Schwarzschild) mass of the system. This one-parameter 
family is the only spherically symmetric solution to 
the vacuum Einstein equations. The spherically symmetric 
gravity can be fully quantized in the minisuperspace 
(frozen) formalism \cite{25,26}. The result of such 
quantization is trivial, quantum functional depends only 
on Schwarzschild mass. Physically it is quite 
understandable. Indeed, one allows the matter sources 
first to collapse classically and then starts to quantize such 
a system. What is left for quantization? Nothing. Mathematically, 
eternal black holes has no dynamical degrees of freedom. 
No real gravitons (because of frozen spherical symmetry), 
no matter source motion. 

To get physically meaningful result we need to introduce 
some dynamical gravitating source. The simplest generalization 
of the point mass is the spherically symmetric 
self-gravitating thin dust shell. The theory of thin shells 
was developed by W.Israel \cite{27} and applied to various problems 
by many authors. For simplicity we consider the 
case when the shell is the only source of gravitational field. 
Then, inside the shell the space-time is flat, and outside it 
is some part of Schwarzschild solution. The dynamics of such 
dust shell is completely described by the single equation \cite{28} 
\begin{equation}
\sqrt{\dot {\rho }^{2}+1}-\sigma \sqrt{\dot {%
\rho }^{2}+1-\frac{2Gm}{\rho }=}\frac{GM}{\rho }
\label{1}
\end{equation}
where $\rho$ is the radius of the shell as a function of proper 
time of an observer sitting on the shell, a dot denotes the 
proper time derivative, $m$ is the total (Schwarzschild) mass 
of the shell, and $M$ is the bare mass (e.g., the sum of the 
masses of constituent dust particles without gravitational mass 
defect). The quantify $\sigma$ is the sign function distinguishing 
two different types of shells. If $\sigma=+1$, the shells moves 
on ``our'' side of the Einstein-Rosen bridge and the radii 
increase when one goes in the outward direction of the shell. 
We will call this the black hole case. If $\sigma=-1$, the shell 
moves beyond the event horizon on the other side or the 
Einstein-Rosen bridge, and radii out of the shell first start to 
decrease, reach the minimal value at the throat and start to 
increase already on our side of the bridge. We will call this 
the wormhole-like case (such a configuration is also called a 
semi-closed world). In what follows we confine ourselves by 
considering the bound motion only. It can be shown that 
\begin{eqnarray}
\frac{m}{M}>\frac{1}{2}\qquad if\qquad \sigma=+1
\\
\frac{m}{M}<\frac{1}{2}\qquad if\qquad \sigma=-1 \nonumber
\label{2}
\end{eqnarray}
The two types of shells can be distinguished by different 
signs of the following inequality ($\rho _{0}$ is the radius 
of the shell at the turning point) 
\begin{eqnarray}
\frac{\partial m}{\partial M}> 0 \qquad if\qquad \sigma=+1
\\
\frac{\partial m}{\partial M}< 0 \qquad if\qquad \sigma=-1 \nonumber
\label{3}
\end{eqnarray}
The seemingly unusual sign in the wormhole case can be easily 
explained. Indeed, the large the bare mass $M$ of the shell, 
the stronger its gravitational field, the more narrow, therefore, 
the throat, and, consequently, the smaller the total mass $m$ 
of the system. 
\newpage
\begin{center}
{\large\bf Quantum Model}
\end{center}
\vskip0.5cm
\noindent
The spherically symmetric space-times with shells can also be 
fully quantized in the minisuperspace formalism \cite{23}. 
All the quantum constraints can be solved, except one. This is 
the Hamiltonian constraint or, Wheeler-DeWitt equation, for 
the shell (here we write it only for the case of bound motion)
\begin{equation}
\Psi (s+i\zeta)+\Psi (s-i\zeta)=\frac{2-\frac{1}{\sqrt{s}}-\frac{M^2}{4 m^2 s}}
{(1-\frac{1}{\sqrt{s}})^{1/2}}\Psi(s)
\label{4}
\end{equation}
Here $s$ is a dimensionless radius squared (normalized by the horizon 
area, $s=R^2/R_g ^2=R^2/4G^2m^2$), $\zeta = \frac{1}{2}(\frac{m_pl}{m})^2$, 
and $i$ is the imaginary unit. The Eqn.(\ref{4}) is an equation in 
finite differences, and the shift in the argument is pure imaginary. 
Thus, the ``good'' solutions should be analytical functions. Besides, 
there are branching points at the horizons (in our case at $s=1$). 
Thus, the wave functions should be analytical on a Riemann's surface 
with a two leaves. The physical reason to consider two Riemann's 
surface is the following. In quantum theory there are no trajectories. 
Thus, even if a shell has parameters $m$ and $M$ (total and bare 
mass) corresponding to the black hole (or wormhole) case, its wave 
function is, in general, everywhere nonzero, ``feel'' both 
infinities on both sides of Einstein-Rosen bridge. The analyticity 
requirement is so stringent that there is no need to solve 
the quantum equation in order to calculate a mass spectrum. One should 
investigate only a behavior of solutions in the vicinity singular 
points (infinities and singularities) and around branching points, 
and then to compare these asymptotics. In such a way the following 
quantum conditions were found for a discrete mass spectrum in the 
case of bound motion \cite{23}.
\begin{eqnarray}
\frac{2  m^2 - M^2}{\sqrt{M^2 - m^2}} = 
\frac{2 m_{pl} ^2}{ m} n
\\
M^2 - m^2 = 2 m_{pl} ^2 (1+2p) \nonumber
\label{5}
\end{eqnarray}
where $n$ and $p$ are integers. The appearance of two quantum 
conditions instead of only one in conventional quantum mechanics is 
due to a nontrivial causal structure of Schwarzschild manifold (two 
infinities!). 

Let us discuss some properties of the spectrum that arises from these 
conditions. 

1. For larger values of quantum number $n$ ($\frac{M^2}{m^2}-1<<1$) one 
can easily derive nonrelativistic Rydberg formula for Kepler's problem, 
$E_{nonrel} = M-m = -\frac{G^2 M^4}{8 n^2}$. 

2. The role of turning point $\rho_0$ is now played by the integer $n$. 
Thus, keeping $n$ constant and calculating 
$\gamma =\frac{\partial m}{\partial M} | _n $ 
one can distinguish 
between a black hole case ($\gamma > 0$) and a wormhole case ($\gamma < 0$). 
It appears that 
$ \frac{\partial m}{\partial M} \vert _n  >0$ for  $n \ge n_0$, 
negative or zero, and 
\begin{equation}
\vert n_0 \vert = E [\sqrt{2} \sqrt{13\sqrt{5}-29} (1 + 2p)]
\label{6}
\end{equation}

3. There exists a minimal possible value for a black hole mass. 
This occurs if $p = n_0 = 0$,
\begin{equation}
m_{min} = \sqrt {2} m_{pl}
\label{7}
\end{equation} 

4. The spectrum described by Eqn.(\ref{5}) is not universal in 
the sense that corresponding wave functions form a 
two-parameter family $\Psi_{n,p}(R)$. 

But for quantum Schwarzschild black hole we expect a one-parameter 
family of wave functions. Quantum black holes should have no 
hairs, otherwise there will be no smooth limit to the classical 
black holes. All this means that our spectrum is not a quantum 
black hole spectrum, and our shell does not collapse (like an 
electron in hydrogen atom). Physically, it is quite understandable, 
because the radiation is yet included into consideration. 

And again, we will use thin shells to model the radiation, but 
this time shells should be null. Let $m_{in}$ and $m_{out}$ be 
a Schwarzschild mass inside  and outside the shell. Then, the 
quantum constraint equation reads as follows \cite{24} 
\begin{equation}
\Psi (m_{in}, m_{out}, s-i\zeta)=
\sqrt{\frac{1-\frac{\mu}{\sqrt{s}}}{1- \frac{1}{\sqrt{s}}}}
\Psi(m_{in}, m_{out},s)
\label{8}
\end{equation}
here $\mu = m_{in} / m_{out}$, $\zeta = \frac{1}{2} m_{pl} ^2 /m_{out} ^2$. 
The existence of the second infinity on the other side of the 
Einstein-Rosen bridge leads to the following quantization condition 
($m = m_{out}$)
\begin{equation}
\delta m = m_{out} - m_{in} = -2m+2 \sqrt{m^2+k m_{pl} ^2},
\label{9}
\end{equation}
where $k$ is an integer. It is interesting to note that if we 
put $k=1$ (minimal radiating energy) and require $\delta m <m$ 
(not more than the total mass can be radiated away), then we 
obtain 
\begin{equation}
m = m_{out} > \frac{2}{\sqrt{5}} m_{pl}.
\label{10}
\end{equation} 
Thus, the black hole with the mass given by  Eqn.(\ref{7}) is not 
radiating and, therefore, it can not be transformed into semi-closed 
world (wormhole-like case). 

The discrete spectrum of radiation (\ref{9}) is universal in the 
sense that it does not depend on the structure and mass spectrum 
of the gravitating source. This means that the energies of 
radiating quanta do not coincide with level spacing of the source. 
The most natural way in resolving such a paradox is to suppose 
that quanta are created in pairs. One of them is radiated away, 
while another one goes inside. Thus, the quantum collapse can not 
proceed without radiating even in the case of spherical symmetry. 
This radiation is accompanying with creation of new shells inside 
the primary shell we started with. We see, that the internal structure 
of quantum black hole is formed during the very process of quantum 
collapse. And if at the beginning we had one shell and knew everything 
about it, then already after the first pulse of radiation we have more 
than one way of creating the inner quantum. So, initially the entropy 
of the system was zero, it starts to grow during the quantum collapse. 
If somehow such a process would stop we would call the resulting 
object ``a quantum black hole''. The natural limit is the transition 
from black hole to the wormhole-like shell. The matter is that such 
a transition requires (at least in quasi-classical regime) insertion 
of an infinitely large volume, and the quasi-classical probability 
for this process is zero. 

Let us write down the spectrum of the shell with nonzero Schwarzschild 
mass, the total mass inside, $m_{in} \ne 0$
\begin{eqnarray}
\frac{2 (\Delta m)^2 - M^2}{\sqrt{M^2 - (\Delta m)^2}} = 
\frac{2 m_{pl} ^2}{\Delta m + m_{in}} n
\\
M^2 - (\Delta m)^2 = 2 (1 +2p) m_{pl} ^2 \nonumber
\label{11}
\end{eqnarray}
Here $\Delta m$ is the total mass of the shell, $M$ is the bare mass, 
the total mass of the system equals $m = m_{out}=\Delta m + m_{in}$. 
For the black hole case $M^2 < 4 m \Delta m$, or 
\begin{equation}
\frac{\Delta m}{M} > 
\frac{1}{2} \left( \sqrt{(\frac{m_{in}}{M})^2 +1} - \frac{m_{in}}{M} \right).
\label{12}
\end{equation}
After switching on the process of radiation governed by Eqn.(\ref{9}), 
the quantum collapse starts. Our computer simulations shows that evolves 
in the ``correct'' direction, e.g. it becomes nearer and nearer to the 
threshold (\ref{12}) between the black hole case and wormhole case. 
The process stops exactly at $n=0$!

The point $n=0$ in the spectrum is very special. Only in such a state 
the shell does not ``feel'' not only the outer regions (what is natural 
for the spherically symmetric configuration) but it does not know anything 
about what is going on inside. It ``feel'' only itself. Such a situation 
reminds the classical (non-spherical) collapse. Finally when all the 
shells (both the primary one and newly produced) are in the 
corresponding states $n_i = 0$, the system does not ``remember'' its 
own history. And this is a quantum black hole. The masses of all the 
shells obey the relation 
\begin{equation}
\Delta m_i = \frac{1}{\sqrt{2}} M_i.
\label{13}
\end{equation}
The subsequent quantum Hawking's evaporation can produced only via 
some collective excitations and formation, e.g., of a long chain of 
microscopic semi-closed worlds.

\newpage
\begin{center}
{\large\bf Classical analog of quantum black hole}
\end{center}
\vskip0.5cm
\noindent

Let us consider large ($m >> m_{pl}$) quantum black holes. The 
number of shells (both primary ones and created during collapse) 
is also very large, and one may hope to construct some classical 
continuous matter distribution that would mimic the properties of 
quantum black holes. First of all, we should translate the 
``no memory'' state ($n=0$ for all the shells) into ``classical 
language''. To do this let us rewrite the Eqn.(\ref{1}) (energy 
constraint equation) for the shell, inside which there is 
some gravitating mass $m_{in}$, 
\begin{equation}
\sqrt{\dot {\rho }^{2}+1-\frac{2Gm_{in}}{\rho}}- \sqrt{\dot {%
\rho }^{2}+1-\frac{2Gm_{out}}{\rho }=}\frac{GM}{\rho }
\label{14}
\end{equation}
and consider a turning point, ($\dot \rho = 0$, $\rho = \rho _0$):
\begin{equation}
\Delta m = m_{out} - m_{in} = 
M \sqrt{1-\frac{2Gm_{in}}{\rho _0}} - \frac{G M^2}{2 \rho _0} .
\label{15}
\end{equation}
It is clear now that in order to make parameters of the shell 
( $\Delta m$ and $M$) not depending on what is going on inside 
we have to put $m_{in} = a \rho _0$.

Our quantum black hole is in a stationary state. Therefore, a 
classical matter distribution should be static. We will consider 
a static perfect fluid with energy density $\varepsilon$ and pressure 
$p$. A static spherically symmetric metric can be written as 
\begin{equation}
d s^2 = e^{\nu}dt^2-e^{\lambda}dr^2-r^2(d\theta^2+\sin^2\theta d \varphi ^2)
\label{16}
\end{equation}
where $\nu$ and $\lambda$ are functions of the radial coordinate 
$r$ only. The relevant Einstein's equations are (prime denotes 
differentiation in $r$)
\begin{eqnarray}
8\pi G\varepsilon = 
- e^{\lambda} (\frac{1}{r^2} - \frac{ \lambda '}{r})+ \frac{1}{r^2},
\nonumber
\\
-8\pi G p = 
- e^{\lambda} (\frac{1}{r^2} - \frac{ \nu '}{r})+ \frac{1}{r^2},
\\
-8\pi G p = 
- \frac{1}{2} e^{\lambda} 
(\nu'' +\frac{\nu'^2}{2} +\frac{\nu'-\lambda'}{r}-\frac{\nu'\lambda'}{2})
 \nonumber
\label{17}
\end{eqnarray}
The first of these equations can be integrated to yield 
\begin{equation}
e^{-\lambda} = 1- \frac{2Gm(r)}{r},
\label{18}
\end{equation}
where
\begin{equation}
m(r)=4\pi \int_0^r \varepsilon r'^2dr'
\label{19}
\end{equation}
is the mass function, that must be identified with $m_{in}$. 
Thus, $m(r)=ar$, and 
\begin{equation}
\varepsilon = \frac{a}{4\pi r^2},\qquad  e^{-\lambda} = 1-2Ga .
\label{20}
\end{equation}
We can also introduce a bare mass function 
\begin{equation}
M(r)=4\pi \int_0^r \varepsilon e^{\frac{\lambda}{2}}r'^2dr' ,
\label{21}
\end{equation}
and from Eqn.(\ref{20}) we get 
\begin{equation}
M(r)=\frac{ar}{\sqrt{1-2Ga}}
\label{22}
\end{equation}
The remaining two equations can now be solved for $p(r)$ and $e^{\nu}$. 
The solution for $p(r)$ that has the correct nonrelativistic limit is 
\begin{equation}
p(r)=\frac{b}{4\pi r^2} ,\qquad 
b=\frac{1}{G}(1-3Ga-\sqrt{1-2Ga}\sqrt{1-4Ga}) ,
\label{23}
\end{equation}
and for $e^{\nu}$ we have 
\begin{equation}
e^{\nu} = C r^{2G \frac{a+b}{1-2Ga}} .
\label{24}
\end{equation}
The constant of integration $C$ can be found from matching of the 
interior and exterior metrics at some boundary $r=r_0$. Let us 
suppose that $r>r_0$ the space-time is empty, so the interior 
should be matched to the Schwarzschild metric. Of course, to 
compensate the jump in pressure ($\Delta p = p(r_0) = p_0$) we 
must introduce some surface tension $\Sigma$. From matching 
conditions it follows that 
\begin{eqnarray}
C = (1-2Ga) r_0 ^{-2G \frac{a+b}{1-2Ga}} ,
\nonumber
\\
e^{\nu} = (1-2Ga)( \frac{r}{r_0}) ^{2G \frac{a+b}{1-2Ga}} ,
\\
\Delta p = \frac{2 \Sigma}{r_0}
 \nonumber
\label{25}
\end{eqnarray}

We would like to stress that the pressure $p$ in our classical 
model is not real  but only effective because it was introduce 
in order to mimic the quantum stationary states. We see, that the 
coefficient $b$ in Eqn.(\ref{23}) becomes a complex number if 
$a>1/4G$. Hence, we must require $a\le 1/4G$, and in the limiting 
point we have the stiffest possible equation of state $\varepsilon=p$ 
It means also that hypothetical quantum collective excitations 
(phonons) would propagate with the speed of light and could be 
considered as massless quasi-particles. It is remarkable that in the 
limiting point we have $m(r)=M(r)/\sqrt{2}$ - the same relation 
as for the total and bare masses in the ``no memory'' state $n=0$! 
The total mass $m_0=m(r_0)$ and the radius $r_0$ in this case are 
related $m_0=4Gr_0$ - twice the horizon size. 

Calculations of Riemann curvature tensor $R^i _{klm}$ and Ricci 
tensor $R_{ik}$ show that if $p<\varepsilon$ ($a\neq b$) there is 
$a$ real singularity at $r=0$. But, surprisingly enough, both 
Riemann and Ricci tensors have finite limits at $r \rightarrow 0$, if 
$\varepsilon =p$ (a=b=1/4G). Therefore we are allowed to introduce 
the so-called topological temperature in the same way as for 
classical black holes. The recipe is the following. One should 
transform the space-time metric by the Wick rotation to the 
Euclidean form and smooth out the canonical singularity by the 
appropriate choice of the period for the imaginary time coordinate. 
The imaginary time coordinate is considered proportional to some 
angle coordinate. In our case the point $r=0$ is already the 
coordinate singularity. The azimuthal angle $\phi$ has the period 
equal to $\pi$. Thus, all other angles should be periodical with 
the period $\pi$. The topological temperature is just the inverse 
of this period. 

The easy exercise shows, that the temperature 
\begin{equation}
T = \frac{1}{2\pi r_0} = \frac{1}{8\pi Gm_0}=T_{BH}
\label{26}
\end{equation}
exactly the same as the Hawking's temperature $T_{BH}$ \cite{5}! 
The very possibility of introducing a temperature provides us 
with the one-parameter family of models with universal distributions 
of energy density and pressure 
\begin{equation}
\varepsilon = p =\frac{1}{16\pi Gr^2},
\label{27}
\end{equation}
the parameter being the total mass $m_0$ or the size $r_0=4Gm_0$.
It should be noted that the two-dimensional part of the metric obtained 
is nothing more but the Rindler's metric, and the null surface $r = 0$ 
serves as the event horizon. 

We can now develop some thermodynamics for our model. First of all 
we should distinguish between global and local thermodynamic 
quantities. The global quantities are those measured by a 
distant observer. He measures the total mass of the system $m_0$ 
and the black temperature $T_{BH} = T_{\infty}$ and does 
not know anything more. Let us assume that this observer is rather 
educated in order to recognize he is dealing with a black hole 
and to write the main thermodynamic relation 
\begin{equation}
dm = TdS.
\label{28}
\end{equation}
In this way he ascribes to a black hole some definite amount of entropy, 
namely, the Hawking-Bekenstein value  
\begin{equation}
S=\frac{1}{4} \frac{(4\pi r_g)^2}{l^2 _{pl}} =
4\pi Gm_0 ^2 = 
4\pi \left (\frac{m_0}{M_{pl}}\right)^2
\label{29}
\end{equation}
Moreover, if this observer is acquainted with, say, the book \cite{29}, 
he can learn from Chapter 3 that, using the Euclidean path integral 
technique, one can calculate a partition function for Schwarzschild 
black hole, 
\begin{equation}
\label{Z}
Z = \sum_n \exp{(- \beta \varepsilon_n)} = 
\exp{\left( - \frac{\beta^2}{16 \pi G} \right)} ,
\end{equation}
with $\beta$ equal to the inverse of Hawking's temperature, 
$\beta = 1/T_H$, and derive the expectation value of the energy, in other 
words, the black hole mass, 
\begin{equation}
\label{E}
m = <E> = - \frac{d}{d \beta} (\ln{Z}) = \frac{\beta}{8 \pi G} .
\end{equation}
Remembering then, that the free energy $F = - T \ln{Z}$ and $F = <E> - T S$, 
he can easily obtain for the entropy $S = \frac{1}{4} \frac{A_h}{l_{Pl}^2}$. 
Calculating the entropy in such a way, the observer builds some statistical 
background for the black hole thermodynamics. However, the obstacle in 
applying usual thermodynamical relations to essentially nonlocal (= global) 
objects, such as black holes, is that the corresponding extensive parameters, 
considered as thermodynamical potentials, are not the homogeneous 
first order functions of all the other extensive parameters. Indeed, 
the entropy $S$ is a quadratic function of the mass (energy), and the free 
energy $F$ is a function of the temperature alone (because there are no 
such extensive parameters like $V$ (volume) and $N$ (``particle'' number) 
which would characterize a black hole thermo-equilibrium state). 

The local observer who measures distributions of energy, pressure 
and local temperature is also rather educated and writes quite 
a different thermodynamic relation 
\begin{equation}
\varepsilon (r)= T(r)s(r)-p(r)-\mu(r) n(r).
\label{30}
\end{equation}
Here $\varepsilon (r)$ and $p(r)$ are energy density and 
pressure, $T(r)$ is the local temperature distribution, $s(r)$ 
is the entropy density, $\mu (r)$ is the chemical potential, 
and $n(r)$ is the number density of some (quasi)''particles''. 
For the energy density and pressure the local observer 
gets, of course, the relation (\ref{27}), and for the 
temperature - the following distribution 
\begin{equation}
T(r) = \frac{1}{\sqrt{2} \pi r},
\label{31}
\end{equation}
which is compatible with the law $T(r) e^{\frac{\nu}{2}} = const$ 
and the boundary condition $T_{\infty} =T_{BH}$. Such 
a distribution is remarkable in that if some outer layer of our 
perfect fluid would be removed, the inner layers would remain in 
thermodynamic equilibrium. And what about the entropy density? 
Surely, the local observer is unable to measure it directly or 
calculate without knowing the microscopic structure of the system, but 
he can receive some information concerning 
the total entropy from the distant observer. This information and the 
measured temperature distribution (\ref{31}) allows him to 
deduce that 
\begin{equation}
s(r)=\frac{1}{8 \sqrt{2} Gr}
\label{32}
\end{equation}
and
\begin{equation}
s(r)T(r)=\frac{1}{16\pi G r^2}
\label{33}
\end{equation}

It is interesting to note that in the main thermodynamic 
equation the contribution from the pressure is compensated 
exactly by the contribution from the temperature and entropy. 
It is noteworthy to remind that the pressure in our classical 
analog model is of quantum mechanical origin as well as the 
black hole temperature. And what is left actually is the 
dust matter we started from in our quantum model, namely, 
\begin{equation}
\varepsilon = \mu n =\frac{1}{16 \pi  Gr^2}
\label{34}
\end{equation}
We may suggest now that the quantum black hole is the 
ensemble of some collective excitations, the black hole 
phonons, and $n(r)$ is just the number density of such 
phonons. 

Knowing equation of state, $\varepsilon = p$, we are able to construct all
the thermodynamical potentials for our system. As an example we show here
how to calculate the energy as a function of the entropy $S$, and the number
particles $N$ (the extensive thermodynamical variables $E$, $S$, $V$,$N$ are 
denoted by capital letters and assumed to have macroscopic but small 
enough values). By the first law of thermodynamics
\begin{equation}
\label{1lt}
dE = TdS - pdV + \mu dN
\end{equation}
where $T=\left .\frac{\partial E}{\partial S} \right |_{V,N}$ is a 
temperature $p= - \left .\frac{\partial E}{\partial V} \right |_{S,N}$ is a
pressure, and $\mu = \left .\frac{\partial E}{\partial N} \right |_{S,V}$ 
is a chemical potential. The energy is additive with respect to the particle
number $N$, hence, $E=Nf(x,y)$ where $x=\frac{S}{N}$ and $y=\frac{N}{V}$.
Since $\varepsilon = \frac{E}{V} = yf(x,y)$ and 
$p = y^2 \frac{\partial f}{\partial y}$ from the equation of state we obtain

\begin{eqnarray}
f = \alpha (x) = n \alpha (x)
\nonumber
\\
\varepsilon = p = n^2 \alpha (x)
\nonumber
\label{f}
\end{eqnarray}
Further,
\begin{eqnarray}
T = n\alpha'(x) 
\nonumber
\\
\mu = n(2 \alpha -x \alpha')
\nonumber
\label{tm}
\end{eqnarray}
But, in any static gravitational field 
$T= T_0/\sqrt{g_{00}}$ and $\mu = \mu_0/ \sqrt{g_{00}}$, so $\mu=\gamma_0 T$,
where $\gamma_0$-some numerical factor. Thus,
\begin{eqnarray}
2 \alpha -x\alpha' = \gamma_0 \alpha',
\nonumber
\\
\alpha(x) = C_0(\gamma_0 + x)^2
\nonumber
\label{x}
\end{eqnarray}
where $C_0$ is a constant of integration. It is easy to see that 
$p/T^2=1/4C_0$. In our specific model $p/T^2 = \pi/8G$, so $C_0=2G/\pi$.
Moreover, because of the relation $\varepsilon=p=Ts=\mu n$ we know that
the free energy $F = E-TS$ is numerically zero. From this we have for 
the entropy
\begin{equation}
\label{ent}
S = \gamma_0 N
\end{equation}
The black hole entropy equals one fourth of the dimensionless horizon area,
and from this we recover the famous Bekenstein-Mukhanov mass spectrum
\begin{equation}
\label{bm}
m=\sqrt{\frac{\gamma_0}{4\pi}} \sqrt{N}m_{pl}
\end{equation}

Note, that our model gives for the free energy an expression quite different 
from that obtained by the use of global thermodynamics. In the latter 
$F = \frac{1}{16 \pi G T}$ which is numerically equal to $\frac{m}{2}$. 
In our case 
\begin{equation}
\label{F}
F = F(T, V, N) = \gamma_0 N T - \frac{V T^2}{4 C_0} ,
\end{equation} 
but the relation (\ref{E}) is nevertheless fulfilled. 

In principle, we can even calculate the remaining unknown coefficient 
$\gamma_0$ using the phonon model. Indeed, since $F=0$, the partition 
function 
\begin{equation}
\label{pf}
Z=\sum_n e^{-\frac{\varepsilon_n}{T}} = 1
\end{equation}
Let us assume that our gravitational phonons have the equidistant energy 
spectrum $\varepsilon_n = \omega n$. Then, $\omega_n - \omega_{n-1} = 
dE = \omega dN$, $(dN = 1)$. Note, that on the static gravitational field 
the ratio $\frac{\omega}{T}$ is an invariant. Therefore, we can use 
$dm$ and $T_{BH}$ (e.i., the increase in the total mass $m$ and the 
Hawking's temperature) instead of local quantities $dE$ and $T$. Then, 
\begin{eqnarray}
\label{omegaz}
\frac{dm}{T_{BH}} = 8 \pi G m dm = dS = \gamma_0 dN = \gamma_0 ,
\\
\\
Z = \sum_n e^{- \frac{\omega n}{T}} = \frac{e^{- \gamma_0}}
{1 - e^{- \omega_0}} = 1 ,
\\
\\
\gamma_0 = \ln{2}
\end{eqnarray}
This just the value advocated by J.Bekenstein and V.Mukhanov in the spirit 
of information theory. If we accept the harmonic oscillator spectrum 
$\varepsilon_n = \omega (n + \frac{1}{2})$ we would obtain 
$\gamma_0 = 2 \ln{\frac{\sqrt{5} + 1}{2}} \approx 1$.
\newpage

\begin{center}
{\Large\bf Acknowlegments}
\end{center}
The author is greatly indebted to the Albert Einstein Institute for kind 
hospitality extended to him. He would like to thank Jurgen Ehlers, 
Kirill Krasnov, Hermann Nicolai, Sergei Odintsov, Alexey Smirnov, 
Thomas Thiemann for helpful discussions. Special thank are to Christine 
Gottschalkson.

\end{document}